\documentclass[aps,prd,twocolumn,nofootinbib,superscriptaddress,longbibliography]{revtex4-2}

\usepackage{amsmath,amssymb,mathtools,bm}
\usepackage{mathrsfs}
\usepackage{graphicx}
\usepackage{hyperref}

\hypersetup{
  colorlinks=true,
  linkcolor=blue,
  citecolor=blue,
  urlcolor=blue
}

\newcommand{\dd}{\mathrm{d}}
\newcommand{\cO}{\mathcal{O}}
\newcommand{\GN}{G_{\mathrm N}}
\newcommand{\E}{E}
\newcommand{\SW}{S_{\mathrm w}}

\begin{document}

\title{An asymptotically conical regular black hole in four-dimensional
nonpolynomial quasitopological gravity}

\author{Jia-Zhou Liu}

\author{Shao-Wen Wei}
\author{Yu-Xiao Liu}
\email{liuyx@lzu.edu.cn}

\affiliation{Lanzhou Center for Theoretical Physics, Key Laboratory of Theoretical Physics of Gansu Province,\\ Key Laboratory of Quantum Theory and Applications of MoE, \\Gansu Provincial Research Center for Basic Disciplines of Quantum Physics, \\ Lanzhou University, Lanzhou 730000, China}

\affiliation{Institute of Theoretical Physics $\&$ Research Center of Gravitation, \\School of Physical Science and Technology, Lanzhou University, Lanzhou 730000, China}

\begin{abstract}
In this work, we construct an asymptotically conical regular black hole in four-dimensional nonpolynomial quasitopological gravity. Starting from a linear choice of the reduced function, which yields an asymptotically conical Schwarzschild-like solution, we introduce a nonlinear deformation that preserves the leading asymptotic cone while producing an exact black hole solution with a de Sitter core and finite curvature invariants. We further establish the corresponding black hole thermodynamics for both solutions within the Hamiltonian and Wald frameworks. The Wald entropy contains rational and logarithmic terms in the regular model, and both solutions satisfy the first law of black hole thermodynamics.
\end{abstract}

\maketitle

\section{Introduction}
\label{sec:introduction}
The Bardeen and Hayward metrics are standard examples of black holes with a regular center~\cite{Bardeen:1968,Hayward:2005gi}. In four-dimensional general relativity, such geometries can be supported by effective matter sources, including nonlinear electrodynamics~\cite{Fan:2016hvf,Liu:2026zzd,PinedoSoto:2026hfm}. Electric and magnetic constructions are subject to consistency conditions~\cite{Bronnikov:2000vy}. Regular centers can also arise from vacuum equations in higher-curvature gravity, as recent explicit constructions demonstrate~\cite{Bueno:2024dgm,Bueno:2025zaj,Borissova:2026wmn}.

Quasitopological gravity is useful here because the static spherical equations reduce to a first integral for a single metric
function~\cite{Myers:2010ru,Oliva:2010eb,Cisterna:2017umf}. Related developments span higher-dimensional cubic, quartic, and
quintic quasitopological gravities, generalized quasitopological theories, and four-dimensional nonpolynomial, nonlocal, and regular
vacuum constructions~\cite{Myers:2010jv,Bueno:2019ltp,Konoplya:2024kih,Horndeski:1974wa,Kobayashi:2011nu,Borissova:2026krh,Carballo-Rubio:2025ntd,Frolov:2024hhe,DiFilippo:2024mwm,Carballo-Rubio:2026mns,Colleaux:2026hat,Bueno:2026oyg,Borissova:2026prd,Dehghani:2011hm,Dehghani:2011vu,Bazrafshan:2012rn,Dehghani:2013ldu,Chernicoff:2016qrc,Dykaar:2017mba,Ghanaatian:2018gdl,Fierro:2020wps,Bazrafshan:2019oan,Bazrafshan:2020fvn,Olamaei:2023csx,Hennigar:2017ego,Tsuda:2026xjc,Frolov:2026rcm,Hao:2025utc,Hennigar:2025yqm}.

The four-dimensional first integral derived in Ref.~\cite{Borissova:2026wmn} involves two functions, $H$ and $G$, but the regular examples analyzed there use $G=0$. We
instead retain a nonzero constant part of $G$. This term contributes to the leading large-radius equation and produces an asymptotically conical vacuum. The geometry near the center is then controlled by the choice of the second function, $H$. For a linear $H$, the resulting solution is a conical Schwarzschild-like geometry with a singular center. A nonlinear choice of $H$ leaves the leading asymptotic cone unchanged but gives an algebraic black hole solution whose center approaches de Sitter space with
finite curvature. Thus, the same asymptotic deficit can be associated with two different types of core: one singular and one regular.
 
Conical Schwarzschild-like geometries and charged, dyonic, and regular extensions in global-monopole and Lorentz-violating models provide related geometric examples~\cite{Barriola:1989hx,Kostelecky:2003fs,
Bluhm:2004ep,Casana:2017jkc,Maluf:2020kgf,Liu:2024axg,Li:2025tcd,Li:2026tae,Li:2026vqn}.  The conical asymptotics are relevant for the physical comparison of these solutions. Although the curvature vanishes at large radius, the solid-angle deficit is a genuine asymptotic datum. It cannot be removed by a coordinate transformation and affects the normalization of the timelike Killing field as well as the definition of the conserved energy. Black hole thermodynamics is sensitive to this asymptotic normalization. Since the asymptotic lapse is not fixed by the cone alone, the physical mass requires a normalized timelike Killing field and a specified energy reference.

The paper  is structured as follows:  Section~\ref{sec:framework} reviews the reduced theory, derives the conditions for asymptotically conical solutions, and constructs the singular and regular solutions. Section~\ref{sec:action} formulates the corresponding covariant actions.
Section~\ref{sec:thermodynamics} derives the Hamiltonian charge, Wald entropy, and first law. Section~\ref{sec:discussion} presents a summary and discussion of
the work.

\section{Four-dimensional nonpolynomial quasitopological gravity}
\label{sec:framework}

We use the covariant spherical reduction introduced in Ref.~\cite{Borissova:2026wmn}. The action is varied on a generic warped-product metric before the static one-function ansatz is imposed. The reduced equations are governed by two functions, $H$ and $G$, and admit a first integral. The sector $G=0$ includes Einstein gravity and the regular black hole solutions constructed in Ref.~\cite{Borissova:2026wmn}. Here we retain a nonzero constant part $G(0)$, which changes the asymptotic vacuum geometry while leaving the higher-curvature terms free to determine the central region. The two
models considered below consequently have the same asymptotic cone but different core geometries.

Consider the four-dimensional warped-product metric
\begin{equation}
	\dd s^2=q_{ab}(y)\dd y^a\dd y^b
	+\varphi(y)^2\dd\Omega_2^2,
	\qquad a,b=0,1 .
	\label{eq:warped-product}
\end{equation}
Here, \(q_{ab}(y)\) is the metric on the two-dimensional spacetime
spanned by the time and radial coordinates, while \(\varphi(y)\) is the areal radius of the two-spheres. This is the general form of a spherically symmetric four-dimensional metric.

The curvature of a warped-product spacetime can be expressed in terms of the two-dimensional curvature and derivatives of \(\varphi\). The following scalar combinations of \(q_{ab}\) and \(\varphi\) are useful in the spherical reduction:
\begin{align}
	\eta&=\frac{\Box\varphi}{\varphi},
	&
	\psi&=\frac{1-(\nabla\varphi)^2}{\varphi^2},
	\nonumber\\
	\tau&=\frac{\nabla_a\nabla_b\varphi
		\nabla^a\nabla^b\varphi}{\varphi^2},
	&
	\mathscr{R}&=R[q].
	\label{eq:spherical-blocks}
\end{align}
In particular, \(\eta\) is the trace of the Hessian of \(\varphi\), whereas \(\tau\) is its squared norm. The quantity \(\psi\) describes the curvature associated with the spherical part of the metric.

The nonpolynomial four-dimensional curvature scalars
\(\mathcal{K}\), \(\mathcal{H}\), \(\mathcal{T}\), and \(\mathcal{P}\)
constructed in Ref.~\cite{Borissova:2026wmn} obey
\begin{equation}
	\left.
	\{\mathcal{K},\mathcal{H},\mathcal{T},\mathcal{P}\}
	\right|_{\mathrm{warped}}
	=
	\{\mathscr{R},\eta,\tau,\psi\}.
	\label{eq:lift-property}
\end{equation}
This relation holds for the warped-product metric~\eqref{eq:warped-product}. The explicit covariant representatives used here are given in Appendix~\ref{app:lift}.

Let \(H_2\), \(H_3\), and \(H_4\) be functions of one variable. Using
the representatives in Eq.~\eqref{eq:lift-property}, we consider the
four-dimensional covariant action~\cite{Borissova:2026wmn}
\begin{align}
	I[g]=&\frac{1}{16\pi\GN}\int\dd^4x\sqrt{-g}\,
	\big[H_2(\mathcal{P})-H_3(\mathcal{P})\mathcal{H}
	+H_4(\mathcal{P})\mathcal{K}
	\nonumber\\
	&+2H_4'(\mathcal{P})
	\left(\mathcal{H}^2-\mathcal{T}\right)
	\big].
	\label{eq:general-action}
\end{align}
A prime denotes differentiation with respect to the argument of the corresponding function. Restricting this action to the warped-product metric and integrating
over the unit two-sphere gives
\begin{align}
	I_{\mathrm{red}}[q,\varphi]
	={}&\frac{1}{4\GN}\int\dd^2y\sqrt{-q}\,\varphi^2
	\big[
	H_2(\psi)-H_3(\psi)\eta
	\nonumber\\
	&+H_4(\psi)\mathscr{R}
	+2H_4'(\psi)\left(\eta^2-\tau\right)
	\big].
	\label{eq:general-reduced-action}
\end{align}
 This reduced action is a two-dimensional Horndeski theory for \(q_{ab}\) and \(\varphi\). In deriving its field equations, we vary the general warped-product fields \(q_{ab}(y)\) and \(\varphi(y)\) before imposing a static ansatz.
We now impose the static, spherically symmetric ansatz
\begin{align}
	\dd s^2={}&-N(r)^2f(r)\dd t^2
	+\frac{\dd r^2}{f(r)}+r^2\dd\Omega_2^2,
	\nonumber\\
	\varphi(r)={}&r .
\end{align}
Since \(q^{rr}=f(r)\), this choice gives\((\nabla\varphi)^2=f(r)\), and hence
\begin{equation}
	\psi(r)=\frac{1-f(r)}{r^2}.
	\label{eq:static-ansatz}
\end{equation}
For this ansatz, the reduced field equations admit a branch with constant lapse, \(N'(r)=0\), provided
\begin{equation}
	H_2'-H_3+2\psi H_3'
	+2H_4-2\psi H_4'
	+4\psi^2H_4''=0 .
	\label{eq:integrability}
\end{equation}
The constant value of \(N\) can be absorbed into a rescaling of the time coordinate, and we therefore set \(N=1\) below. Here the functions and their derivatives are evaluated at \(\psi\).
This condition ensures that the static spherical field equations admit a sector with a constant lapse, so that only one independent metric function remains. It is the condition derived for this class of reduced actions in Ref.~\cite{Borissova:2026wmn}.

It is useful to combine the three functions \(H_i\) into two functions:
\begin{align}
	H(\psi)&=H_2(\psi)+\psi H_3(\psi),
	\nonumber\\
	G(\psi)&=H_3(\psi)-4H_4(\psi)
	+4\psi H_4'(\psi).
	\label{eq:HG-definitions}
\end{align}

The remaining equation has the first integral
\begin{equation}
	r(1+2f)G(\psi)+r^3H(\psi)=12M,
	\label{eq:master-equation}
\end{equation}
where \(M\) is an integration constant with dimensions of length~\cite{Borissova:2026wmn}. This first integral is algebraic in \(f(r)\)
at each radius, since \(\psi=(1-f)/r^2\). The relation between \(M\)
and the physical mass depends on the normalization of the asymptotic
timelike Killing vector; that normalization will be fixed below using
the covariant charge.

As a check, Einstein--Hilbert gravity is recovered by choosing
\begin{equation}
	H_2(\psi)=2\psi,\qquad H_3(\psi)=4,\qquad H_4(\psi)=1.
	\label{eq:Einstein-His}
\end{equation}
These functions satisfy Eq.~\eqref{eq:integrability} and give
\begin{equation}
	H(\psi)=6\psi,\qquad G(\psi)=0.
\end{equation}
Equation~\eqref{eq:master-equation} then yields \(f(r)=1-2M/r\), the Schwarzschild solution.

To determine the leading asymptotic behavior, assume that \(H\) and \(G\) are regular near \(\psi=0\) and have the expansions
\begin{align}
	H(\psi)&=H'(0)\psi
	+\cO\!\left(|\psi|^{1+\epsilon}\right),
	\qquad \epsilon>\frac12,
	\nonumber\\
	G(\psi)&=G(0)+\cO(|\psi|).
	\label{eq:HG-IR}
\end{align}
For an asymptotically conical metric, \(f(r)\) approaches a constant, so \(\psi=\cO(r^{-2})\). The nonlinear correction to \(H\) therefore contributes to the first integral at order
\[
r^3\cO\!\left(|\psi|^{1+\epsilon}\right)
=\cO\!\left(r^{1-2\epsilon}\right)=o(1).
\]
Consequently, this nonlinear part does not affect the terms of order \(r\) or \(r^0\) in the asymptotic field equation. These leading terms are determined only by \(H'(0)\) and \(G(0)\).

Let the large-radius behavior of the metric function be
\begin{equation}
	f(r)=c-\frac{2m}{r}+o(r^{-1}),
	\qquad c>0.
	\label{eq:f-asymptotic}
\end{equation}
Then
\begin{equation}
	\psi(r)=\frac{1-c}{r^2}
	+\frac{2m}{r^3}+o(r^{-3}).
	\label{eq:psi-asymptotic}
\end{equation}
The term of order $r$ in Eq.~\eqref{eq:master-equation} vanishes if
\begin{equation}
	G(0)(1+2c)+H'(0)(1-c)=0.
	\label{eq:asymptotic-c-condition}
\end{equation}
The constant term is
\begin{equation}
	2m\big[H'(0)-2G(0)\big]=12M.
\end{equation}
For $H'(0)-2G(0)\neq0$,
\begin{equation}
	c=\frac{H'(0)+G(0)}{H'(0)-2G(0)},
	\qquad
	m=\frac{6M}{H'(0)-2G(0)}.
	\label{eq:general-ca}
\end{equation}

Since \(f(r)=c+\mathcal{O}(r^{-1})\) as \(r\to\infty\), the metric with
the constant lapse fixed to \(N=1\) has the asymptotic form
\begin{equation}
	\dd s^2\simeq -c\,\dd t^2+\frac{\dd r^2}{c}
	+r^2\dd\Omega_2^2,
	\qquad c>0.
\end{equation}
Introducing the asymptotically normalized time and proper radial
coordinate,
\begin{equation}
	T=\sqrt{c}\,t,\qquad
	\rho=\frac{r}{\sqrt{c}},
\end{equation}
we obtain
\begin{equation}
	\dd s^2\simeq
	-\dd T^2+\dd\rho^2+c\,\rho^2\dd\Omega_2^2.
	\label{eq:asymptotic-conical-metric}
\end{equation}
A sphere at proper radius $\rho$ has area $4\pi c\rho^2$. The solid
angle at infinity is $4\pi c$, with deficit
\begin{equation}
	\Delta\Omega=4\pi(1-c).
	\label{eq:solid-angle-deficit}
\end{equation}

Choose
\begin{equation}
	H'(0)=6,\qquad G(0)=\gamma,
\end{equation}
and parametrize the asymptotic constant as
\begin{equation}
	c=\frac{1}{1+\ell}.
	\label{eq:c-ell}
\end{equation}
Equation~\eqref{eq:general-ca} yields
\begin{equation}
	\gamma=-\frac{6\ell}{\ell+3}.
	\label{eq:gamma-ell}
\end{equation}
The solid-angle-deficit branch, including the asymptotically flat limit,
is
\begin{equation}
	\ell\geq0,
	\qquad
	-6<\gamma\leq0,
	\qquad
	0<c\leq1.
	\label{eq:physical-coupling-range}
\end{equation}
The interval $-1<\ell<0$ gives $c>1$ and corresponds to a solid-angle
excess. The following analysis focuses on $\ell\geq0$.

Two choices of $H$ at fixed $G=\gamma$ are considered below.

\subsection{Conical Schwarzschild-like solution}

Consider the linear choice
\begin{equation}
	H(\psi)=6\psi,\qquad G(\psi)=\gamma.
	\label{eq:singular-HG}
\end{equation}
The master equation becomes
\begin{equation}
	r\left[\gamma(1+2f)+6(1-f)\right]=12M,
	\label{eq:linear-master}
\end{equation}
with exact solution
\begin{equation}
	f(r)=c-\frac{2m}{r},
	\qquad
	c=\frac{6+\gamma}{6-2\gamma},
	\qquad
	M=m\left(1-\frac{\gamma}{3}\right).
	\label{eq:exact-f}
\end{equation}

Although the constant lapse was set to one in deriving the master equation, the time coordinate relevant for the asymptotic charges is
normalized by requiring \(g_{tt}\to-1\) as \(r\to\infty\). Since \(f(r)\to c\), this fixes
\begin{equation}
	N^2c=1,
	\qquad
	N^2=\frac{1}{c}=1+\ell .
	\label{eq:time-normalization}
\end{equation}
The metric in this normalization is
\begin{equation}
	\begin{aligned}
		\dd s^2={}&-(1+\ell)
		\left(\frac{1}{1+\ell}-\frac{2m}{r}\right)\dd t^2
		\\
		&+\frac{\dd r^2}{
			\frac{1}{1+\ell}-\frac{2m}{r}}
		+r^2\dd\Omega_2^2 .
	\end{aligned}
	\label{eq:physical-metric}
\end{equation}
This solution has the asymptotic conical geometry and solid-angle deficit given in Eqs.~\eqref{eq:asymptotic-conical-metric} and~\eqref{eq:solid-angle-deficit}.

A constant rescaling of $t$ leaves scalar curvature invariants
unchanged. For $f=c-2m/r$, they are
\begin{align}
	R&=\frac{2(1-c)}{r^2},
	\label{eq:Ricci-scalar}\\
	R_{\mu\nu\rho\sigma}R^{\mu\nu\rho\sigma}
	&=\frac{48m^2}{r^6}
	+\frac{16m(1-c)}{r^5}
	+\frac{4(1-c)^2}{r^4}.
	\label{eq:Kretschmann}
\end{align}
The curvature vanishes at large radius but diverges at $r=0$, so even
the $m=0$ conical vacuum is singular at the apex when $c\neq1$.

For $m>0$ and $c>0$, the event horizon is located at
\begin{equation}
	r_h=\frac{2m}{c}=2m(1+\ell).
	\label{eq:horizon-radius}
\end{equation}
The solution has one nondegenerate horizon and a Schwarzschild-like
central singularity.

\subsection{Regular conical Hayward solution}
\label{sec:regular-solution}

For the regular model, we choose a rational deformation of the linear
function while keeping \(G\) constant:
\begin{equation}
	H(\psi)=\frac{6\psi}{1-L^2\psi},
	\qquad
	G(\psi)=\gamma,
	\label{eq:regular-HG}
\end{equation}
where \(L>0\) is a length scale. This choice has the same leading small-$\psi$ behavior as the linear model but differs from it at quadratic and higher orders in $\psi$.

Using \(f=1-r^2\psi\), so that \(1+2f=3-2r^2\psi\), the master equation takes the form
\begin{equation}
	\gamma r(3-2r^2\psi)
	+\frac{6r^3\psi}{1-L^2\psi}=12M .
	\label{eq:regular-master-psi}
\end{equation}
The equation is understood away from the pole
\(1-L^2\psi=0\). Multiplying by \(1-L^2\psi\) and collecting powers of
\(\psi\) gives
\begin{equation}
	A_2\psi^2+A_1\psi+A_0=0,
	\label{eq:regular-quadratic}
\end{equation}
where
\begin{align}
	A_2&=2\gamma L^2r^3,\nonumber\\
	A_1&=12ML^2-3\gamma L^2r+(6-2\gamma)r^3,\nonumber\\
	A_0&=3\gamma r-12M.
	\label{eq:regular-coefficients}
\end{align}
For \(\gamma\neq0\), this is a genuine quadratic equation in \(\psi\). When \(\gamma=0\), the quadratic coefficient vanishes and the equation reduces to the linear \(G=0\) branch, which will also be obtained as the
continuous \(\gamma\to0\) limit.
The discriminant is
\begin{align}
	\Delta&=A_1^2-4A_2A_0\nonumber\\
	&=\left[L^2(12M-3\gamma r)
	+2(3+\gamma)r^3\right]^2-48\gamma r^6 .
	\label{eq:regular-discriminant}
\end{align}
When \(M>0\) and \(-6<\gamma<0\), \(\Delta\) is positive for every \(r>0\), and Eq.~\eqref{eq:regular-quadratic} has two roots,
\begin{equation}
	\psi_\pm(r)=\frac{-A_1\pm\sqrt{\Delta}}{2A_2}.
\end{equation}

We select the root with the plus sign,
\begin{equation}
	\psi(r)=\frac{-A_1+\sqrt{\Delta}}{2A_2}.
	\label{eq:regular-root-quadratic}
\end{equation}
Using
$\Delta=A_1^2-4A_2A_0$, the same root can be written as
\begin{equation}
	\psi(r)
	=-\frac{2A_0}{A_1+\sqrt{\Delta}}.
	\label{eq:regular-root}
\end{equation}
This latter form is preferable for taking the limit
$A_2\propto\gamma\to0$.

Substituting Eq.~\eqref{eq:regular-root} into
$f=1-r^2\psi$ gives
\begin{equation}
	f(r)=1-
	\frac{6r^2(4M-\gamma r)}
	{L^2(12M-3\gamma r)+(6-2\gamma)r^3+\sqrt{\Delta}}.
	\label{eq:regular-f-exact}
\end{equation}
This expression contains no explicit inverse power of $\gamma$ and
therefore has a smooth $\gamma\to0$ limit.

Indeed, when $\gamma=0$, the quadratic equation degenerates into the linear equation
\begin{equation}
(12ML^2+6r^3)\psi-12M=0.
\end{equation}
Hence
\begin{equation}
	\psi(r)=\frac{2M}{r^3+2ML^2},
	\qquad
	f(r)=1-\frac{2Mr^2}{r^3+2ML^2}.
	\label{eq:hayward-limit}
\end{equation}
This is the standard Hayward metric. In the present construction, however, it arises as a vacuum solution of the modified gravitational theory. Expanding the solution for \(r\to\infty\) gives

\begin{equation}
	f(r)=c-\frac{2m}{r}+\frac{b}{r^2}
	+\mathcal{O}(r^{-3}),
	\qquad
	c=\frac{1}{1+\ell},
	\label{eq:regular-asymptotic}
\end{equation}
where
\begin{equation}
	m=\frac{6M}{6-2\gamma},
	\qquad
	b=\frac{6L^2(1-c)^2}{6-2\gamma}
	=\frac{L^2\ell^2(\ell+3)}
	{3(\ell+1)^3}.
	\label{eq:regular-b}
\end{equation}
Thus, at fixed $\gamma$ (or $\ell$), the leading asymptotic cone and the $1/r$ term are independent of the regularization scale $L$. For $\gamma\neq0$, the first $L$-dependent correction appears at order $r^{-2}$. In the asymptotically flat limit $\gamma=0$, however, $b=0$; for the Hayward solution the first $L$-dependent term appears at order $r^{-4}$.

Near the center, the branch with $M>0$ behaves as
\begin{equation}
	f(r)=1-\frac{r^2}{L^2}
	+\frac{r^5}{2ML^4}+\mathcal{O}(r^6),
	\qquad r\to0.
	\label{eq:regular-center-expansion}
\end{equation}
 Accordingly, the curvature invariants at the center are
\begin{align}
	R(0)&=\frac{12}{L^2},\\
	R_{\mu\nu\rho\sigma}R^{\mu\nu\rho\sigma}(0)
	&=\frac{24}{L^4},\\
	C_{\mu\nu\rho\sigma}C^{\mu\nu\rho\sigma}(0)&=0.
\end{align}
Hence the central curvature is finite, and the Weyl curvature vanishes at the center.

The horizon condition at $r=r_h$ is

\begin{equation}
	M(r_h)=\frac{r_h}{12}
	\left[\gamma+\frac{6r_h^2}{r_h^2-L^2}\right].
	\label{eq:regular-horizon-mass}
\end{equation}
For \(M>0\) and \(-6<\gamma\leq0\), this equation has no solution with \(0<r_h<L\), because its right-hand side is negative in that interval. Thus, physical horizons satisfy \(r_h>L\).

Introducing\[y=\frac{r_h^2}{L^2}>1,\] the derivative of the  mass is
\begin{equation}
	\frac{\dd M}{\dd r_h}
	=
	\frac{L^4 P(y)}
	{12(r_h^2-L^2)^2},
	P(y)=\gamma(y-1)^2+6y(y-3).
	\label{eq:regular-mass-derivative}
\end{equation}
An extremal horizon is a degenerate zero of the metric function, satisfying
\[
f(r_h)=0,\qquad f'(r_h)=0.
\]
Using \(f(r_h)=0\), we express the mass in terms of the horizon radius as
\(M=M(r_h)\), and differentiating this relation shows that \(f'(r_h)=0\) is equivalent to $\dd M /\dd r_h=0.$
Therefore, the extremality condition reduces to \(P(y)=0\). For \(-6<\gamma\leq0\) the two roots are
\begin{equation}y_\pm=\frac{\gamma+9\pm\sqrt{81+12\gamma}}{\gamma+6}.\end{equation}

Only the root with the plus sign lies in the physical domain \(y>1\). Thus,
\begin{equation}
y_{\rm ext}
=
\frac{\gamma+9+\sqrt{81+12\gamma}}{\gamma+6}>1.
\end{equation} The corresponding mass is
\begin{equation}
	M_{\rm ext}
	=\frac{L\,y_{\rm ext}^{3/2}}
	{(y_{\rm ext}-1)^2}.
\end{equation}

For \(M>0\), the horizon mass is positive only for \(r_h>L\). Moreover,
\begin{equation}
	M_h(r_h)\to+\infty
	\quad (r_h\to L^+),
	\qquad
	M_h(r_h)\to+\infty
 (r_h\to\infty).
	\label{eq:regular-mass-limits}
\end{equation}
Since \(y_{\rm ext}\) is the only stationary point in this interval, it is the global minimum of \(M_h(r_h)\). Therefore,
\begin{equation}
\begin{cases}
	M>M_{\rm ext}:& \text{two horizons},\\
	M=M_{\rm ext}:& \text{one degenerate horizon},\\
	0<M<M_{\rm ext}:& \text{no horizon}.
\end{cases}
\end{equation}

Differentiating the master equation and evaluating it at a horizon gives
\begin{equation}
	f'(r_h)=
	\frac{\gamma(r_h^2-L^2)^2
		+6r_h^2(r_h^2-3L^2)}
	{r_h\left[6r_h^4
		-2\gamma(r_h^2-L^2)^2\right]} .
	\label{eq:regular-fprime}
\end{equation}
The numerator is \(L^4P(y)\). For \(r_h>L\) and \(\gamma\leq0\), the denominator is positive, so the sign of \(f'(r_h)\) is the sign of \(P(y)\). Thus, on the two-horizon branch,
\[
f'(r_{\rm in})<0,\qquad f'(r_{\rm out})>0,
\] whereas \(f'(r_h)=0\) at the extremal horizon.

With the asymptotically normalized time coordinate, \(N=c^{-1/2}\), the temperature of the outer horizon is
\begin{equation}
	T_{\rm H}=\frac{N f'(r_{\rm out})}{4\pi},
\end{equation}
and it approaches zero as \(M\to M_{\rm ext}\).
The solution has the limiting behavior
\begin{equation}
	f(r)=
	\begin{cases}
		1-r^2/L^2+o(r^2),&r\to0,\\[1mm]
		c-2m/r+o(r^{-1}),&r\to\infty.
	\end{cases}
	\label{eq:regular-two-limits}
\end{equation}
The center approaches de Sitter space with curvature scale $L$, whereas
the curvature at infinity falls as
\begin{equation}
	R=\frac{2(1-c)}{r^2}+o(r^{-2}).
\end{equation}
The solid-angle deficit is therefore an asymptotic property and does not
fix the geometry of the core.
\section{Reduced actions and covariant representatives}
\label{sec:action}

The master equation is supplemented by functions $H_i$ satisfying the integrability condition. For each choice of $H$ and $G$, the reduction prescription gives a corresponding covariant representative. The two models below differ in the nonlinear part of $H(\psi)$ but share the same leading asymptotic data.
\subsection{Action for the conical Schwarzschild-like solution}

For the conical Schwarzschild-like model, one convenient solution of the
integrability condition is
\begin{equation}
	H_2(\psi)=\left(2+\frac{\gamma}{3}\right)\psi,
	H_3(\psi)=4-\frac{\gamma}{3},
	H_4(\psi)=1-\frac{\gamma}{3}.
	\label{eq:reconstructed-His}
\end{equation}
These functions satisfy
\begin{equation}
	H_2+\psi H_3=6\psi,
	\qquad
	H_3-4H_4+4\psi H_4'=\gamma .
	\label{eq:HG-check}
\end{equation}
A term proportional to $\psi$ may be added to $H_4$ without changing $H$ or $G$. It shifts the Wald entropy by an $r_h$-independent constant,
which will be set to zero.

Because $H_4$ is constant, the last term in
Eq.~\eqref{eq:general-action} vanishes. The four-dimensional Ricci scalar
reduces as
\begin{equation}
	\left.R\right|_{\mathrm{warped}}
	=\mathscr{R}-4\eta+2\psi.
	\label{eq:R-reduction}
\end{equation}
With
$\mathcal{P}|_{\mathrm{warped}}=\psi$ and
$\mathcal{H}|_{\mathrm{warped}}=\eta$, the action takes the form
\begin{equation}
	I_\gamma[g]=\frac{1}{16\pi\GN}\int\dd^4x\sqrt{-g}
	\left[
	\left(1-\frac{\gamma}{3}\right)R
	+\gamma\left(\mathcal{P}-\mathcal{H}\right)
	\right].
	\label{eq:compact-action-gamma}
\end{equation}
On the warped-product metric,
\begin{align}
	&\left[
	\left(1-\frac{\gamma}{3}\right)R
	+\gamma(\mathcal{P}-\mathcal{H})
	\right]_{\mathrm{warped}}
	\nonumber\\
	&\qquad=
	\left(1-\frac{\gamma}{3}\right)\mathscr{R}
	+\left(2+\frac{\gamma}{3}\right)\psi
	-\left(4-\frac{\gamma}{3}\right)\eta .
	\label{eq:reduced-compact-action}
\end{align}
The reduced expression agrees with Eq.~\eqref{eq:reconstructed-His}, and the integrability condition becomes
\begin{equation}
	\left(2+\frac{\gamma}{3}\right)
	-\left(4-\frac{\gamma}{3}\right)
	+2\left(1-\frac{\gamma}{3}\right)=0.
	\label{eq:integrability-check}
\end{equation}

On the domain where the covariant representatives are defined, the principle of symmetric criticality identifies the reduced equations with the spherical restriction of the four-dimensional covariant equations~\cite{Palais:1979rca,Fels:2001rv}. Hence, the metric in Eq.~\eqref{eq:physical-metric} solves the corresponding covariant vacuum equations within this spherical sector.

Using \(\gamma=-6\ell/(\ell+3)\), the same action can be written in
terms of the deficit parameter \(\ell\) as
\begin{align}
	I_\ell[g]=&\frac{1}{16\pi\GN}\int\dd^4x\sqrt{-g}\Bigg[
	\frac{3(1+\ell)}{\ell+3}R
	\nonumber\\
	&-\frac{6\ell}{\ell+3}
	\left(\mathcal{P}-\mathcal{H}\right)
	\Bigg].
	\label{eq:compact-action-ell}
\end{align}
At \(\ell=0\), the coefficient of \(\mathcal{P}-\mathcal{H}\) vanishes, and the action reduces exactly to the Einstein--Hilbert action. For \(\ell\neq0\), the nonpolynomial dependence on the curvature is encoded in the covariant scalars \(\mathcal{P}\) and \(\mathcal{H}\). Their explicit representatives, whose spherical reductions are \(\psi\) and \(\eta\), respectively, are given in Appendix~\ref{app:lift}.

\subsection{Action for the regular conical Hayward solution}
\label{sec:regular-action}

A convenient reconstruction of the regular model is
\begin{align}
	H_4(\psi)&=1-\frac{L^2\psi}{1-L^2\psi}
	+2L^2\psi\ln\!\left[
	\frac{\psi_0(1-L^2\psi)}{\psi}
	\right]
	-\frac{\gamma}{3},
	\nonumber\\
	H_3(\psi)&=\frac{4}{(1-L^2\psi)^2}
	-\frac{\gamma}{3},
	\nonumber\\
	H_2(\psi)&=\frac{2\psi(1-3L^2\psi)}
	{(1-L^2\psi)^2}
	+\frac{\gamma}{3}\psi .
	\label{eq:regular-reconstructed-His}
\end{align}
Here \(\psi_0>0\) is an arbitrary inverse-length-squared scale that makes
the logarithm dimensionless. The
real branch is
\begin{equation}
	0<\psi<L^{-2},
	\qquad
	0<\mathcal{P}<L^{-2}
\end{equation}
for the covariant representative. A change of $\psi_0$ shifts the Wald
entropy by an $r_h$-independent constant.

The functions in Eq.~\eqref{eq:regular-reconstructed-His} reproduce the
functions specified in Eq.~\eqref{eq:regular-HG}:
\begin{equation}
	H_2+\psi H_3=\frac{6\psi}{1-L^2\psi},
	\qquad
	H_3-4H_4+4\psi H_4'=\gamma .
	\label{eq:regular-HG-check}
\end{equation}
Substitution into Eq.~\eqref{eq:general-action}, together with the covariant representatives given in Appendix~\ref{app:lift}, therefore
defines the corresponding covariant action. At fixed \(\psi\), the limit
\(L\to0\) reduces \(H(\psi)\) to \(6\psi\), so the reduced action approaches the constant-\(G\) model.

The selected branch approaches the endpoint \(\psi=L^{-2}\) as \(r\to0\). This endpoint does not signal a curvature
singularity: the individual reconstruction functions \(H_i(\psi)\) may diverge there, while the metric and its curvature invariants remain finite, as shown in Sec.~\ref{sec:regular-solution}. To examine the behavior of the reduced action, define
\begin{equation}
	\delta=1-L^2\psi .
\end{equation}
For \(M>0\), the central expansion gives
\begin{equation}
	\delta=\frac{r^3}{2ML^2}+\mathcal O(r^4),
	\qquad r\to0.
\end{equation}
After this expansion is substituted into the reduced Lagrangian, the terms proportional to \(\delta^{-2}\) and \(\delta^{-1}\) cancel. The
remaining angularly integrated on-shell radial density is
\begin{equation}
	\mathcal L_{\rm sph}^{\rm on\text{-}shell}
	=\frac{Nr^2}{4G_{\rm N}L^2}
	\left[
	12\ln\!\left(\frac{\psi_0r^3}{2M}\right)
	+\mathcal O(1)
	\right]
	=\mathcal O(r^2\ln r).
	\label{eq:regular-action-integrability}
\end{equation}
Since
\[
\int_0^\varepsilon r^2\ln r\,\dd r
\]
is finite, the on-shell reduced action is locally integrable at the center.

\section{Thermodynamics}
\label{sec:thermodynamics}
We do not consider extended black hole thermodynamics, in which the couplings are varied as thermodynamic variables. Thus, all action parameters are held fixed. In the linear model, $\gamma$ is fixed, whereas both $\gamma$ and $L$ are fixed in the regular model. Since the asymptotic constant $c$ depends only on $\gamma$, the asymptotic cone remains unchanged as the mass parameter $M$ varies.
We normalize the time coordinate by requiring $g_{tt}\to-1$ at infinity, which gives $N=c^{-1/2}$. The lapse $N$ is therefore also held fixed in the thermodynamic variations.

The first integral~\eqref{eq:master-equation} is independent of the
constant lapse. We define the radial mass function ~\cite{Borissova:2026rbi}
\begin{equation}
	\Omega(r,f)
	=
	\frac{1}{12\GN}
	\left[
	r(1+2f)G(\psi)+r^3H(\psi)
	\right].
	\label{eq:thermo-generator}
\end{equation}
The field equation can then be written as
\begin{equation}
	\Omega(r,f(r))=\frac{M}{\GN}.
	\label{eq:onshell-omega}
\end{equation}
Thus, $\Omega$ is constant along a given vacuum solution. The physical
energy is obtained from the covariant charge associated with the
properly normalized Killing field.

In the Iyer--Wald formalism, the variation of the Hamiltonian generated by \(\xi=\partial_t\) is obtained by integrating the surface-charge form
\begin{equation}
	k_\xi=\delta Q_\xi-\xi\mathbin{\cdot}\Theta,
\end{equation}
where $Q_\xi$ is the Noether charge and $\Theta$ is the presymplectic
potential~\cite{Wald:1993nt,Iyer:1994ys}. The variation is first
performed for the general warped-product fields $q_{ab}$ and
$\varphi$, and the static ansatz is imposed only afterwards.
At fixed radius, $\varphi=r$ implies
\[
\delta\varphi=0,
\qquad
\delta\psi=-\frac{\delta f}{r^2}.
\]
Evaluating the reduced covariant charge on the static solution gives
\begin{align}
	\int_{S_r}Q_{\partial_t}
	&=
	\frac{N}{4\GN}
	\left[
	r^2H_4 f'+rfG
	\right],
	\nonumber\\
	\int_{S_r}\partial_t\mathbin{\cdot}\Theta
	&=
	\frac{N}{4\GN}
	\bigg\{
	r^2H_4\,\delta f'
	+
	\left[
	rG+2r(H_4-\psi H_4')-H_4'f'
	\right]\delta f
	\bigg\}.
	\label{eq:general-charge-ingredients}
\end{align}
Here, $f'$ denotes differentiation with respect to $r$, while the primes on $H_i$, $H$, and $G$ denote differentiation with respect to $\psi$. The orientation is chosen such that the Schwarzschild energy is positive.

Varying the first line of Eq.~\eqref{eq:general-charge-ingredients} and subtracting the second one removes the terms proportional to $\delta f'$. We obtain
\begin{equation}
	\int_{S_r}k_{\partial_t}
	=
	-\frac{N}{4\GN}
	\left[
	2r(H_4-\psi H_4')
	+\frac{f}{r}G'
	\right]\delta f.
	\label{eq:reduced-surface-charge}
\end{equation}
The integrability condition together with the definitions of $H$ and
$G$ gives
\begin{equation}
	H'-2G+\psi G'
	=
	6(H_4-\psi H_4').
	\label{eq:thermo-integrability-identity}
\end{equation}
Using this identity in Eq.~\eqref{eq:reduced-surface-charge}, we find
\begin{align}
	\int_{S_r}k_{\partial_t}
	&=
	N\,\partial_f\Omega(r,f)\,\delta f
	=
	N\,\delta\Omega(r,f),
	\nonumber\\
	\partial_f\Omega
	&=
	\frac{1}{12\GN}
	\left[
	2rG-\frac{1+2f}{r}G'-rH'
	\right].
	\label{eq:general-direct-charge}
\end{align}
The last equality follows because the variation is taken at fixed $r$
and with all couplings held fixed.

For a Killing background, with the perturbation satisfying the
linearized vacuum equations, the surface charge obeys
\[
\dd k_{\partial_t}=0.
\]
Its integral is therefore independent of the sphere on which it is
evaluated. At infinity, Eq.~\eqref{eq:onshell-omega} gives
\begin{equation}
	\delta H_{\partial_t}
	=
	\frac{N}{\GN}\,\delta M.
	\label{eq:hamiltonian-energy}
\end{equation}
Since $N$ is fixed along the family, this variation is integrable. We
choose the zero of energy such that the solution with $M=0$ has zero
energy:
\begin{equation}
	\E=\frac{NM}{\GN}.
	\label{eq:physical-energy}
\end{equation}

We next compute the entropy at a nonextremal outer horizon $r=r_h$.
At the horizon,
\[
f(r_h)=0,
\qquad
\psi(r_h)=r_h^{-2},
\]
and the temperature associated with the normalized Killing field is
\begin{equation}
	T=\frac{Nf'(r_h)}{4\pi}.
	\label{eq:general-temperature}
\end{equation}
On the bifurcation sphere $\mathcal B$, the Noether charge satisfies
\begin{equation}
		\int_{\mathcal B}Q_{\partial_t}
	=
	\frac{\kappa}{2\pi}\SW
	=
	T\SW,
\end{equation}
where
\[
\kappa=\frac{Nf'(r_h)}{2}.
\]
The term proportional to $rfG$ in Eq.~\eqref{eq:general-charge-ingredients}
vanishes at the horizon. Therefore,
\begin{equation}
	\SW
	=
	\frac{\pi r_h^2}{\GN}H_4(r_h^{-2}).
	\label{eq:general-Wald-entropy}
\end{equation}
This is the Wald entropy obtained from the coefficient of
$\mathscr R$ in the reduced action. Its additive constant is fixed by
the chosen covariant representative. In particular, adding a term
proportional to $\psi$ to $H_4$ leaves $H$ and $G$ unchanged but shifts
the entropy by an $r_h$-independent constant.

Differentiating Eq.~\eqref{eq:general-Wald-entropy} and using
Eq.~\eqref{eq:thermo-integrability-identity}, we obtain
\begin{equation}
	\frac{\dd\SW}{\dd r_h}
	=
	\frac{2\pi r_h}{\GN}
	\left[H_4-\psi H_4'\right]_{\psi=r_h^{-2}}
	=
	-4\pi\,\partial_f\Omega(r_h,0).
	\label{eq:entropy-generator}
\end{equation}

Along a vacuum solution, $\Omega$ is constant:
\begin{equation}
	0
	=
	\frac{\dd}{\dd r}\Omega(r,f(r))
	=
	\partial_r\Omega+f'\partial_f\Omega.
	\label{eq:radial-mass-conservation}
\end{equation}
On the horizon curve,
\[
\E(r_h)=N\Omega(r_h,0).
\]
Since the couplings and $N$ are fixed,
\[
\frac{\dd\E}{\dd r_h}
=
N\,\partial_r\Omega(r_h,0).
\]
Combining this relation with
Eq.~\eqref{eq:radial-mass-conservation}, the temperature, and
Eq.~\eqref{eq:entropy-generator}, gives
\begin{equation}
	T\frac{\dd\SW}{\dd r_h}
	=
	-Nf'(r_h)\partial_f\Omega(r_h,0)
	=
	N\partial_r\Omega(r_h,0)
	=
	\frac{\dd\E}{\dd r_h}.
\end{equation}
Hence the first law is
\begin{equation}
	\dd\E=T\,\dd\SW.
	\label{eq:general-first-law}
\end{equation}

\subsection{Conical Schwarzschild-like solution}
We now specialize to the linear conical Schwarzschild-like model,
\begin{equation}
	H(\psi)=6\psi,
	\qquad
	G(\psi)=\gamma.
\end{equation}
The charge variation reduces to
\begin{equation}
	\int_{S_r}k_{\partial_t}
	=
	-\frac{	(1-\gamma/3)Nr}{2\GN}\,\delta f(r).
\end{equation}
Since $f(r)=c-2m/r$ and $M=	(1-\gamma/3)m$, this gives
\begin{equation}
	\delta H_{\partial_t}
	=
	\frac{	(1-\gamma/3)N}{\GN}\,\delta m
	=
	\frac{N}{\GN}\,\delta M,
\end{equation}
in agreement with Eq.~\eqref{eq:hamiltonian-energy}.

The Wald entropy is
\begin{equation}
	\SW
	=
	(1-\gamma/3)\,\frac{\pi r_h^2}{\GN}
	=
	\frac{3(1+\ell)}{3+\ell}
	\frac{A_h}{4\GN},
	\qquad
	A_h=4\pi r_h^2.
	\label{eq:singular-Wald-entropy}
\end{equation}
Here
\begin{equation}
	\gamma=-\frac{6\ell}{\ell+3},
	\qquad
	c=\frac{1}{1+\ell},
	\qquad
	N=\sqrt{1+\ell}.
\end{equation}
Using the horizon radius $r_h=2m/c$, the thermodynamic quantities
become
\begin{align}
	\E
	&=
	\frac{3\sqrt{1+\ell}}
	{2\GN(\ell+3)}\,r_h,
	\nonumber\\
	T
	&=
	\frac{1}{4\pi r_h\sqrt{1+\ell}},
	\nonumber\\
	\SW
	&=
	\frac{3\pi(1+\ell)}
	{\GN(\ell+3)}\,r_h^2.
\end{align}
At fixed $\ell$, these expressions satisfy
\begin{equation}
	\dd\E=T\,\dd\SW,
	\qquad
	\E=2T\SW.
	\label{eq:first-law-Smarr}
\end{equation}
The deviation from the Einstein area law is determined by the gravitational action.

\subsection{Regular conical Hayward solution}
On the regular branch, 
\begin{equation}
	H(\psi)=\frac{6\psi}{1-L^2\psi},
	\qquad
	G(\psi)=\gamma.
\end{equation}
Equation~\eqref{eq:general-direct-charge} gives
\begin{equation}
	\int_{S_r}k_{\partial_t}
	=
	\frac{Nr}{12\GN}
	\left[
	2\gamma-\frac{6}{(1-L^2\psi(r))^2}
	\right]\delta f(r).
	\label{eq:regular-direct-charge}
\end{equation}
On shell, this is $N\delta M/\GN$. We therefore use $\E=NM/\GN$ with the same energy reference as in the linear model. To display the thermodynamic quantities explicitly, define
\begin{align}
	D_h
	&=
	r_h^2-L^2,
	\qquad
	Q_h
	=
	3r_h^4-\gamma D_h^2,
	\nonumber\\
	\mathcal N_h
	&=
	\gamma D_h^2
	+6r_h^2(r_h^2-3L^2).
	\label{eq:regular-thermo-polynomials}
\end{align}
When $r_h>L$ and $-6<\gamma\leq0$, both $D_h$ and $Q_h$ are positive.
The horizon equation gives
\begin{equation}
	M
	=
	\frac{r_h}{12}
	\left[
	\gamma+\frac{6r_h^2}{D_h}
	\right].
	\label{eq:regular-horizon-mass-thermo}
\end{equation}
Therefore,
\begin{align}
	\E
	&=
	\frac{Nr_h}{12\GN}
	\left[
	\gamma+\frac{6r_h^2}{D_h}
	\right],
	\nonumber\\
	T
	&=
	\frac{N\mathcal N_h}
	{8\pi r_hQ_h}.
	\label{eq:regular-energy-temperature}
\end{align}
Substituting the reconstructed $H_4$ from Eq.~\eqref{eq:regular-reconstructed-His} into Eq.~\eqref{eq:general-Wald-entropy} gives
\begin{equation}
\SW	=
	\frac{\pi}{\GN}
	\left[
	\left(1-\frac{\gamma}{3}\right)r_h^2
	-\frac{L^2r_h^2}{D_h}
	+2L^2\ln(\psi_0D_h)
	\right].
	\label{eq:regular-entropy}
\end{equation}
The auxiliary scale $\psi_0$ changes only the additive constant in the entropy. Differentiating the mass and entropy gives
\begin{equation}
	\frac{\dd M}{\dd r_h}
	=
	\frac{\mathcal N_h}{12D_h^2},
	\qquad
	\frac{\dd \SW}{\dd r_h}
	=
	\frac{2\pi r_hQ_h}{3\GN D_h^2}.
	\label{eq:regular-thermo-derivatives}
\end{equation}
Together with $\E=NM/\GN$, these relations imply
\begin{equation}
	\dd\E
	=
	T\,\dd \SW.
\end{equation}
The extremal horizon is determined by
\begin{equation}
	\mathcal N_h=0.
\end{equation}
At extremality,
\begin{equation}
	T=0,
	\qquad
	\frac{\dd\E}{\dd r_h}=0.
\end{equation}
On the outer-horizon branch, one has
\begin{equation}
	f'(r_h)>0,
	\qquad
	\frac{\dd \SW}{\dd r_h}>0.
\end{equation}

\section{Conclusion}
\label{sec:discussion}
Building on the four-dimensional construction of Ref.~\cite{Borissova:2026wmn}, we have extended the analysis by allowing a nonzero constant part of the reduced function, $G(\psi)=\gamma$. We find that the constant part of $G$ fixes the asymptotic conical geometry, whereas the nonlinear completion of $H$ determines the behavior of the core. 

We construct two asymptotically conical black hole solutions. The first is a conical Schwarzschild-like solution: it is asymptotically conical at large radius, but its curvature diverges at the central singularity. The second solution is a conical regular black hole. It has the same type of asymptotic conical geometry, while the curvature remains finite and the center is regular with de Sitter-like behavior. Its horizon structure depends on the mass: the solution can have two horizons, an extremal horizon, or no horizon. Hence an asymptotic cone alone does not determine whether the core is singular or regular.

For a fixed choice of the couplings and a fixed normalization of the asymptotic Killing field, the Hamiltonian energy and the Wald entropy obey the first law. The regular branch has an entropy containing rational and logarithmic terms. These contributions originate from the nonminimal coupling function $H_4$ in the reconstructed action and should therefore be interpreted as higher-curvature gravitational effects.

Although the conical Schwarzschild-like branch is singular at the center, a suitable nonlinear completion of the reduced dynamics yields a regular black hole with the same asymptotic conical geometry. This construction is not unique. Other choices of the reduced functions may give rise to additional asymptotically conical black holes, whose geometric and thermodynamic properties remain to be explored.
\section*{Acknowledgments}
This work was supported by the National Natural Science Foundation of China (Grants No. 12475056, No. 12475055, No. 12247101), the 111 Project (Grant No. B20063), the Fundamental Research
Funds for the Central Universities 
(Grants No. lzujbky-2025-jdzx07), 
the Natural Science
Foundation of Gansu Province (No. 
22JR5RA389 and
No. 25JRRA799), and Gansu Province's Top Leading Talent Support Plan.

\appendix

\section{Explicit covariant representatives}
\label{app:lift}

The curvature invariants are
\begin{align}
	\mathcal I_{C^2}
	&=C_{\mu\nu\rho\sigma}C^{\mu\nu\rho\sigma},
	\label{eq:IC2}\\
	\mathcal I_{C^3}
	&=C_{\mu\nu}{}^{\rho\sigma}
	C_{\rho\sigma}{}^{\alpha\beta}
	C_{\alpha\beta}{}^{\mu\nu},
	\label{eq:IC3}\\
	\mathcal I_{\widehat R^2}
	&=\widehat R_\mu{}^\nu\widehat R_\nu{}^\mu,
	\label{eq:IRhat2}\\
	\mathcal I_{\widehat R^3}
	&=\widehat R_\mu{}^\nu\widehat R_\nu{}^\rho
	\widehat R_\rho{}^\mu,
	\label{eq:IRhat3}\\
	\mathcal I_{R^2C}
	&=R^{\mu\nu}R^{\rho\sigma}C_{\rho\mu\nu\sigma},
	\label{eq:IR2C}
\end{align}
where  the traceless Ricci tensor is
\begin{equation}
	\widehat R_{\mu\nu}
	=R_{\mu\nu}-\frac14Rg_{\mu\nu}.
	\label{eq:traceless-Ricci}
\end{equation}
The common denominator appearing below is
\begin{equation}
	\mathscr D=
	\mathcal I_{C^2}\mathcal I_{R^2C}
	+2\mathcal I_{\widehat R^2}\mathcal I_{C^3}.
	\label{eq:lift-common-denominator}
\end{equation}

One set of covariant representatives constructed in
Ref.~\cite{Borissova:2026wmn} is
\begin{equation}
	\mathcal H
	=-\frac16R+\frac{\mathcal I_{C^3}}{\mathcal I_{C^2}},
	\label{eq:covariant-H}
\end{equation}
\begin{align}
	\mathcal T={}&\frac{R^2}{72}
	+\frac{\mathcal I_{\widehat R^2}}{6}
	-\frac{R\mathcal I_{C^3}}
	{6\mathcal I_{C^2}}
	+\frac{\mathcal I_{C^3}^2}
	{2\mathcal I_{C^2}^2}
	\nonumber\\
	&+\frac{\mathcal I_{C^2}\mathcal I_{R^2C}}
	{12\mathcal I_{C^3}},
	\label{eq:covariant-T}
\end{align}
\begin{align}
	\mathcal P={}&\frac{1}{12}R
	+\frac{\mathcal I_{C^3}}{\mathcal I_{C^2}}
	\nonumber\\
	&-\frac{\mathcal I_{\widehat R^3}\mathcal I_{C^3}}
	{\mathscr D},
	\label{eq:covariant-P}
\end{align}
and
\begin{align}
	\mathcal K={}&\frac{1}{\mathscr D}\Bigg[
	\frac16R\mathcal I_{C^2}\mathcal I_{R^2C}
	+\frac13R\mathcal I_{\widehat R^2}\mathcal I_{C^3}
	\nonumber\\
	&+4\frac{\mathcal I_{\widehat R^2}\mathcal I_{C^3}^2}
	{\mathcal I_{C^2}}
	+2\mathcal I_{\widehat R^3}\mathcal I_{C^3}
	+2\mathcal I_{C^3}\mathcal I_{R^2C}
	\Bigg].
	\label{eq:covariant-K}
\end{align}
These expressions are understood away from zeros of the displayed
denominators and extended to the spherical configurations by continuity.

On the warped-product metrics~\eqref{eq:warped-product}, they satisfy
\begin{equation}
	\left.
	\{\mathcal K,\mathcal H,\mathcal T,\mathcal P\}
	\right|_{\mathrm{warped}}
	=
	\{\mathscr R,\eta,\tau,\psi\}.
	\label{eq:KHTP-reduction}
\end{equation}
Combining the functions reconstructed in Sec.~\ref{sec:action} with Eqs.~\eqref{eq:covariant-H}--\eqref{eq:covariant-K} gives the four-dimensional representatives used in Eq.~\eqref{eq:general-action}. Their variation follows the generic-warped-product prescription of Sec.~\ref{sec:framework}.

For the regular solution, apparent $0/0$ expressions are defined through
the spherical limit. Near the center,
\begin{align}
	\mathcal I_{C^2}&=\cO(r^6),
	&
	\mathcal I_{C^3}&=\cO(r^9),
	\nonumber\\
	\mathcal I_{\widehat R^2}&=\cO(r^6),
	&
	\mathcal I_{\widehat R^3}&=\cO(r^9),
	\nonumber\\
	\mathcal I_{R^2C}&=\cO(r^9),
	&
	\mathscr D&=\cO(r^{15}).
	\label{eq:regular-invariant-orders}
\end{align}
The ratios then behave as
\begin{equation}
	\frac{\mathcal I_{C^3}}{\mathcal I_{C^2}}
	=\cO(r^3),
	\qquad
	\frac{\mathcal I_{\widehat R^3}\mathcal I_{C^3}}
	{\mathscr D}
	=\cO(r^3).
\end{equation}
The leading terms in the numerator and denominator of $\mathcal K$ have the same order. Its spherical limit, as well as those of $\mathcal H$, $\mathcal T$, and $\mathcal P$, is finite and agrees with Eq.~\eqref{eq:KHTP-reduction}.

\bibliographystyle{apsrev4-2}
\bibliography{conical_bh_nqtg}

@inproceedings{Bardeen:1968,
    author = "Bardeen, J.",
    title = "{Non-singular general relativistic gravitational collapse}",
    booktitle = "{Proceedings of the 5th International Conference on Gravitation and the Theory of Relativity}",
    month = "9",
    pages = "87",
    year = "1968"
}

@article{Hayward:2005gi,
    author = "Hayward, Sean A.",
    title = "{Formation and evaporation of regular black holes}",
    eprint = "gr-qc/0506126",
    archivePrefix = "arXiv",
    doi = "10.1103/PhysRevLett.96.031103",
    journal = "Phys. Rev. Lett.",
    volume = "96",
    pages = "031103",
    year = "2006"
}

@article{Fan:2016hvf,
    author = "Fan, Zhong-Ying and Wang, Xiaobao",
    title = "{Construction of Regular Black Holes in General Relativity}",
    eprint = "1610.02636",
    archivePrefix = "arXiv",
    primaryClass = "gr-qc",
    doi = "10.1103/PhysRevD.94.124027",
    journal = "Phys. Rev. D",
    volume = "94",
    number = "12",
    pages = "124027",
    year = "2016"
}

@article{Bronnikov:2000vy,
    author = "Bronnikov, Kirill A.",
    title = "{Regular magnetic black holes and monopoles from nonlinear electrodynamics}",
    eprint = "gr-qc/0006014",
    archivePrefix = "arXiv",
    doi = "10.1103/PhysRevD.63.044005",
    journal = "Phys. Rev. D",
    volume = "63",
    pages = "044005",
    year = "2001"
}

@article{Myers:2010ru,
    author = "Myers, Robert C. and Robinson, Brandon",
    title = "{Black Holes in Quasi-topological Gravity}",
    eprint = "1003.5357",
    archivePrefix = "arXiv",
    primaryClass = "gr-qc",
    doi = "10.1007/JHEP08(2010)067",
    journal = "JHEP",
    volume = "08",
    pages = "067",
    year = "2010"
}

@article{Oliva:2010eb,
    author = "Oliva, Julio and Ray, Sourya",
    title = "{A new cubic theory of gravity in five dimensions: Black hole, Birkhoff's theorem and C-function}",
    eprint = "1003.4773",
    archivePrefix = "arXiv",
    primaryClass = "gr-qc",
    reportNumber = "CECS-PHY-10-03",
    doi = "10.1088/0264-9381/27/22/225002",
    journal = "Class. Quant. Grav.",
    volume = "27",
    pages = "225002",
    year = "2010"
}

@article{Cisterna:2017umf,
    author = "Cisterna, Adolfo and Guajardo, Luis and Hassaine, Mokhtar and Oliva, Julio",
    title = "{Quintic quasi-topological gravity}",
    eprint = "1702.04676",
    archivePrefix = "arXiv",
    primaryClass = "hep-th",
    doi = "10.1007/JHEP04(2017)066",
    journal = "JHEP",
    volume = "04",
    pages = "066",
    year = "2017"
}

@article{Myers:2010jv,
    author = "Myers, Robert C. and Paulos, Miguel F. and Sinha, Aninda",
    title = "{Holographic studies of quasi-topological gravity}",
    eprint = "1004.2055",
    archivePrefix = "arXiv",
    primaryClass = "hep-th",
    reportNumber = "DAMTP-2010-28",
    doi = "10.1007/JHEP08(2010)035",
    journal = "JHEP",
    volume = "08",
    pages = "035",
    year = "2010"
}

@article{Bueno:2019ltp,
    author = "Bueno, Pablo and Cano, Pablo A. and Moreno, Javier and Murcia, {\'A}ngel",
    title = "{All higher-curvature gravities as Generalized quasi-topological gravities}",
    eprint = "1906.00987",
    archivePrefix = "arXiv",
    primaryClass = "hep-th",
    doi = "10.1007/JHEP11(2019)062",
    journal = "JHEP",
    volume = "11",
    pages = "062",
    year = "2019"
}

@article{Bueno:2024dgm,
    author = "Bueno, Pablo and Cano, Pablo A. and Hennigar, Robie A.",
    title = "{Regular black holes from pure gravity}",
    eprint = "2403.04827",
    archivePrefix = "arXiv",
    primaryClass = "gr-qc",
    doi = "10.1016/j.physletb.2025.139260",
    journal = "Phys. Lett. B",
    volume = "861",
    pages = "139260",
    year = "2025"
}

@article{Konoplya:2024kih,
    author = "Konoplya, R. A. and Zhidenko, A.",
    title = "{Dymnikova black hole from an infinite tower of higher-curvature corrections}",
    eprint = "2404.09063",
    archivePrefix = "arXiv",
    primaryClass = "gr-qc",
    doi = "10.1016/j.physletb.2024.138945",
    journal = "Phys. Lett. B",
    volume = "856",
    pages = "138945",
    year = "2024"
}

@article{Horndeski:1974wa,
    author = "Horndeski, Gregory Walter",
    title = "{Second-order scalar-tensor field equations in a four-dimensional space}",
    doi = "10.1007/BF01807638",
    journal = "Int. J. Theor. Phys.",
    volume = "10",
    pages = "363--384",
    year = "1974"
}

@article{Kobayashi:2011nu,
    author = "Kobayashi, Tsutomu and Yamaguchi, Masahide and Yokoyama, Jun'ichi",
    title = "{Generalized G-inflation: Inflation with the most general second-order field equations}",
    eprint = "1105.5723",
    archivePrefix = "arXiv",
    primaryClass = "hep-th",
    reportNumber = "KUNS-2339, RESCEU-9-11",
    doi = "10.1143/PTP.126.511",
    journal = "Prog. Theor. Phys.",
    volume = "126",
    pages = "511--529",
    year = "2011"
}

@article{Borissova:2026krh,
    author = "Borissova, Johanna",
    title = "{All 2D generalized dilaton theories from d{\ensuremath{\geq}}4 gravities}",
    eprint = "2603.06786",
    archivePrefix = "arXiv",
    primaryClass = "hep-th",
    reportNumber = "Imperial/TP/2026/JB/03",
    doi = "10.1103/v3j1-j6jg",
    journal = "Phys. Rev. D",
    volume = "113",
    number = "12",
    pages = "124088",
    year = "2026"
}

@article{Carballo-Rubio:2025ntd,
    author = "Carballo-Rubio, Ra{\'u}l",
    title = "{Master field equations for spherically symmetric gravitational fields beyond general relativity}",
    eprint = "2507.15920",
    archivePrefix = "arXiv",
    primaryClass = "gr-qc",
    doi = "10.1038/s41467-026-69035-6",
    journal = "Nature Commun.",
    volume = "17",
    number = "1",
    pages = "1399",
    year = "2026"
}

@article{Borissova:2026wmn,
    author = "Borissova, Johanna and Carballo-Rubio, Ra{\'u}l",
    title = "{Regular black holes from pure gravity in four dimensions}",
    eprint = "2602.16773",
    archivePrefix = "arXiv",
    primaryClass = "gr-qc",
    reportNumber = "Imperial/TP/2026/JB/02",
    doi = "10.1103/6x2z-qbkh",
    journal = "Phys. Rev. D",
    volume = "113",
    number = "12",
    pages = "124004",
    year = "2026"
}

@article{Frolov:2024hhe,
    author = "Frolov, Valeri P. and Koek, Alex and Soto, Jose Pinedo and Zelnikov, Andrei",
    title = "{Regular black holes inspired by quasitopological gravity}",
    eprint = "2411.16050",
    archivePrefix = "arXiv",
    primaryClass = "gr-qc",
    reportNumber = "Alberta Thy 11-24",
    doi = "10.1103/PhysRevD.111.044034",
    journal = "Phys. Rev. D",
    volume = "111",
    number = "4",
    pages = "044034",
    year = "2025"
}

@article{DiFilippo:2024mwm,
    author = "Di Filippo, Francesco and Kol{\'a}{\v{r}}, Ivan and Kubiznak, David",
    title = "{Inner-extremal regular black holes from pure gravity}",
    eprint = "2404.07058",
    archivePrefix = "arXiv",
    primaryClass = "gr-qc",
    doi = "10.1103/PhysRevD.111.L041505",
    journal = "Phys. Rev. D",
    volume = "111",
    number = "4",
    pages = "L041505",
    year = "2025"
}

@article{Barriola:1989hx,
    author = "Barriola, Manuel and Vilenkin, Alexander",
    title = "{Gravitational Field of a Global Monopole}",
    reportNumber = "TUTP-89-4",
    doi = "10.1103/PhysRevLett.63.341",
    journal = "Phys. Rev. Lett.",
    volume = "63",
    pages = "341",
    year = "1989"
}

@article{Kostelecky:2003fs,
    author = "Kostelecky, V. Alan",
    title = "{Gravity, Lorentz violation, and the standard model}",
    eprint = "hep-th/0312310",
    archivePrefix = "arXiv",
    reportNumber = "IUHET-461",
    doi = "10.1103/PhysRevD.69.105009",
    journal = "Phys. Rev. D",
    volume = "69",
    pages = "105009",
    year = "2004"
}

@article{Bluhm:2004ep,
    author = "Bluhm, Robert and Kostelecky, V. Alan",
    title = "{Spontaneous Lorentz violation, Nambu-Goldstone modes, and gravity}",
    eprint = "hep-th/0412320",
    archivePrefix = "arXiv",
    reportNumber = "IUHET-478",
    doi = "10.1103/PhysRevD.71.065008",
    journal = "Phys. Rev. D",
    volume = "71",
    pages = "065008",
    year = "2005"
}

@article{Casana:2017jkc,
    author = "Casana, R. and Cavalcante, A. and Poulis, F. P. and Santos, E. B.",
    title = "{Exact Schwarzschild-like solution in a bumblebee gravity model}",
    eprint = "1711.02273",
    archivePrefix = "arXiv",
    primaryClass = "gr-qc",
    doi = "10.1103/PhysRevD.97.104001",
    journal = "Phys. Rev. D",
    volume = "97",
    number = "10",
    pages = "104001",
    year = "2018"
}

@article{Maluf:2020kgf,
    author = "Maluf, R. V. and Neves, Juliano C. S.",
    title = "{Black holes with a cosmological constant in bumblebee gravity}",
    eprint = "2011.12841",
    archivePrefix = "arXiv",
    primaryClass = "gr-qc",
    doi = "10.1103/PhysRevD.103.044002",
    journal = "Phys. Rev. D",
    volume = "103",
    number = "4",
    pages = "044002",
    year = "2021"
}

@article{Liu:2024axg,
    author = "Liu, Jia-Zhou and Guo, Wen-Di and Wei, Shao-Wen and Liu, Yu-Xiao",
    title = "{Charged spherically symmetric and slowly rotating charged black hole solutions in bumblebee gravity}",
    eprint = "2407.08396",
    archivePrefix = "arXiv",
    primaryClass = "gr-qc",
    doi = "10.1140/epjc/s10052-025-13859-x",
    journal = "Eur. Phys. J. C",
    volume = "85",
    number = "2",
    pages = "145",
    year = "2025"
}

@article{Li:2025tcd,
    author = "Li, Shoulong and Liang, Liang and Ma, Liang",
    title = "{Dyonic RN-like and Taub-NUT-like black holes in Einstein-bumblebee gravity}",
    eprint = "2510.04405",
    archivePrefix = "arXiv",
    primaryClass = "gr-qc",
    doi = "10.1088/1475-7516/2026/03/005",
    journal = "JCAP",
    volume = "03",
    pages = "005",
    year = "2026"
}

@article{Li:2026tae,
    author = "Li, Zhi-Chao and Lu, H.",
    title = "{When Bumblebee Meets NLED: Lorentz-Violating Black Holes and Regular Spacetimes}",
    eprint = "2605.13963",
    archivePrefix = "arXiv",
    primaryClass = "gr-qc",
    journal = "Eur. Phys. J. C",
    volume = "86",
    pages = "959",
    year = "2026"
}

@misc{Li:2026vqn,
    author = "Li, Zhi-Chao and Lu, H.",
    title = "{Lorentz-Violating (Regular) Black Holes in Einstein Gravity}",
    eprint = "2605.27615",
    archivePrefix = "arXiv",
    primaryClass = "gr-qc",
    month = "5",
    year = "2026"
}

@article{Borissova:2026rbi,
    author = "Borissova, Johanna",
    title = "{gttgrr=--1 black hole thermodynamics in general quasitopological gravity}",
    eprint = "2604.24101",
    archivePrefix = "arXiv",
    primaryClass = "gr-qc",
    reportNumber = "Imperial/TP/2026/JB/05",
    doi = "10.1103/gbdk-xm4c",
    journal = "Phys. Rev. D",
    volume = "114",
    number = "6",
    pages = "064038",
    year = "2026"
}

@article{Wald:1993nt,
    author = "Wald, Robert M.",
    title = "{Black hole entropy is the Noether charge}",
    eprint = "gr-qc/9307038",
    archivePrefix = "arXiv",
    reportNumber = "EFI-93-42",
    doi = "10.1103/PhysRevD.48.R3427",
    journal = "Phys. Rev. D",
    volume = "48",
    number = "8",
    pages = "R3427--R3431",
    year = "1993"
}

@article{Iyer:1994ys,
    author = "Iyer, Vivek and Wald, Robert M.",
    title = "{Some properties of Noether charge and a proposal for dynamical black hole entropy}",
    eprint = "gr-qc/9403028",
    archivePrefix = "arXiv",
    doi = "10.1103/PhysRevD.50.846",
    journal = "Phys. Rev. D",
    volume = "50",
    pages = "846--864",
    year = "1994"
}

@article{Palais:1979rca,
    author = "Palais, Richard S.",
    title = "{The principle of symmetric criticality}",
    doi = "10.1007/BF01941322",
    journal = "Commun. Math. Phys.",
    volume = "69",
    number = "1",
    pages = "19--30",
    year = "1979"
}

@article{Fels:2001rv,
    author = "Fels, Mark E. and Torre, Charles G.",
    title = "{The Principle of symmetric criticality in general relativity}",
    eprint = "gr-qc/0108033",
    archivePrefix = "arXiv",
    doi = "10.1088/0264-9381/19/4/303",
    journal = "Class. Quant. Grav.",
    volume = "19",
    pages = "641--676",
    year = "2002"
}

@misc{Liu:2026zzd,
    author = "Liu, Hong-Lin and Feng, Zhong-Wen and Jiang, Qing-Quan and Zhou, Xia and Mu, Xue-Ling",
    title = "{Globally regular charged black holes in non-polynomial quasi-topological gravity with Born-Infeld electrodynamics}",
    eprint = "2609.07576",
    archivePrefix = "arXiv",
    primaryClass = "gr-qc",
    month = "9",
    year = "2026"
}

@article{PinedoSoto:2026hfm,
    author = "Pinedo Soto, Jose and Frolov, Valeri P.",
    title = "{Charged black holes in quasitopological gravity coupled to Born-Infeld nonlinear electrodynamics}",
    eprint = "2604.06632",
    archivePrefix = "arXiv",
    primaryClass = "gr-qc",
    doi = "10.1103/8px4-2pyk",
    journal = "Phys. Rev. D",
    volume = "113",
    number = "12",
    pages = "124044",
    year = "2026"
}

@misc{Carballo-Rubio:2026mns,
    author = "Carballo-Rubio, Ra{\'u}l and Mazza, Jacopo",
    title = "{Exploring the construction of field equations for regular black holes with dynamical regularization scales}",
    eprint = "2609.01733",
    archivePrefix = "arXiv",
    primaryClass = "gr-qc",
    month = "9",
    year = "2026"
}

@misc{Colleaux:2026hat,
    author = "Coll{\'e}aux, Aimeric",
    title = "{Rational regular black holes in non-polynomial gravity}",
    eprint = "2608.17158",
    archivePrefix = "arXiv",
    primaryClass = "gr-qc",
    month = "8",
    year = "2026"
}

@misc{Bueno:2026oyg,
    author = "Bueno, Pablo and Cano, Pablo A. and Hennigar, Robie A. and Murcia, {\'A}ngel J.",
    title = "{Regular Black Holes in Nonlocal Quasitopological Gravity}",
    eprint = "2607.07790",
    archivePrefix = "arXiv",
    primaryClass = "gr-qc",
    month = "7",
    year = "2026"
}

@misc{Borissova:2026prd,
    author = "Borissova, Johanna",
    title = "{Formation of extremal regular black holes}",
    eprint = "2608.02908",
    archivePrefix = "arXiv",
    primaryClass = "gr-qc",
    reportNumber = "Imperial/TP/2026/JB/07",
    month = "8",
    year = "2026"
}

@article{Dehghani:2011hm,
    author = "Dehghani, M. H. and Vahidinia, M. H.",
    title = "{Surface Terms of Quasitopological Gravity and Thermodynamics of Charged Rotating Black Branes}",
    eprint = "1108.4235",
    archivePrefix = "arXiv",
    primaryClass = "hep-th",
    doi = "10.1103/PhysRevD.84.084044",
    journal = "Phys. Rev. D",
    volume = "84",
    pages = "084044",
    year = "2011"
}

@article{Dehghani:2011vu,
    author = "Dehghani, M. H. and Bazrafshan, A. and Mann, R. B. and Mehdizadeh, M. R. and Ghanaatian, M. and Vahidinia, M. H.",
    title = "{Black Holes in Quartic Quasitopological Gravity}",
    eprint = "1109.4708",
    archivePrefix = "arXiv",
    primaryClass = "hep-th",
    doi = "10.1103/PhysRevD.85.104009",
    journal = "Phys. Rev. D",
    volume = "85",
    pages = "104009",
    year = "2012"
}

@article{Bazrafshan:2012rn,
    author = "Bazrafshan, A. and Dehghani, M. H. and Ghanaatian, M.",
    title = "{Surface Terms of Quartic Quasitopological Gravity and Thermodynamics of Nonlinear Charged Rotating Black Branes}",
    eprint = "1209.0246",
    archivePrefix = "arXiv",
    primaryClass = "hep-th",
    doi = "10.1103/PhysRevD.86.104043",
    journal = "Phys. Rev. D",
    volume = "86",
    pages = "104043",
    year = "2012"
}

@article{Dehghani:2013ldu,
    author = "Dehghani, M. H. and Vahidinia, M. H.",
    title = "{Quartic Quasi-topological Gravity, Black Holes and Holography}",
    eprint = "1307.0330",
    archivePrefix = "arXiv",
    primaryClass = "hep-th",
    doi = "10.1007/JHEP10(2013)210",
    journal = "JHEP",
    volume = "10",
    pages = "210",
    year = "2013"
}

@article{Chernicoff:2016qrc,
    author = "Chernicoff, Mariano and Fierro, Octavio and Giribet, Gaston and Oliva, Julio",
    title = "{Black holes in quasi-topological gravity and conformal couplings}",
    eprint = "1612.00389",
    archivePrefix = "arXiv",
    primaryClass = "hep-th",
    doi = "10.1007/JHEP02(2017)010",
    journal = "JHEP",
    volume = "02",
    pages = "010",
    year = "2017"
}

@article{Dykaar:2017mba,
    author = "Dykaar, Hannah and Hennigar, Robie A. and Mann, Robert B.",
    title = "{Hairy black holes in cubic quasi-topological gravity}",
    eprint = "1703.01633",
    archivePrefix = "arXiv",
    primaryClass = "hep-th",
    doi = "10.1007/JHEP05(2017)045",
    journal = "JHEP",
    volume = "05",
    pages = "045",
    year = "2017"
}

@article{Hennigar:2017ego,
    author = "Hennigar, Robie A. and Kubiz{\v{n}}{\'a}k, David and Mann, Robert B.",
    title = "{Generalized quasitopological gravity}",
    eprint = "1703.01631",
    archivePrefix = "arXiv",
    primaryClass = "hep-th",
    doi = "10.1103/PhysRevD.95.104042",
    journal = "Phys. Rev. D",
    volume = "95",
    number = "10",
    pages = "104042",
    year = "2017"
}

@article{Ghanaatian:2018gdl,
    author = "Ghanaatian, M. and Naeimipour, F. and Bazrafshan, A. and Abkar, M.",
    title = "{Charged black holes in quartic quasi-topological gravity}",
    eprint = "1801.05692",
    archivePrefix = "arXiv",
    primaryClass = "gr-qc",
    doi = "10.1103/PhysRevD.97.104054",
    journal = "Phys. Rev. D",
    volume = "97",
    number = "10",
    pages = "104054",
    year = "2018"
}

@article{Bazrafshan:2019oan,
    author = "Bazrafshan, A. and Naeimipour, F. and Olamaei, A. R. and Ghanaatian, M.",
    title = "{Thermodynamics of static solutions in ($n$+1)-dimensional Quintic Quasitopological gravity}",
    eprint = "1910.03051",
    archivePrefix = "arXiv",
    primaryClass = "hep-th",
    doi = "10.1103/PhysRevD.100.124018",
    journal = "Phys. Rev. D",
    volume = "100",
    number = "12",
    pages = "124018",
    year = "2019"
}

@article{Fierro:2020wps,
    author = "Fierro, Octavio and Mora, Nicolas and Oliva, Julio",
    title = "{Slowly rotating black holes in quasitopological gravity}",
    eprint = "2012.06618",
    archivePrefix = "arXiv",
    primaryClass = "hep-th",
    doi = "10.1103/PhysRevD.103.064004",
    journal = "Phys. Rev. D",
    volume = "103",
    number = "6",
    pages = "064004",
    year = "2021"
}

@article{Bazrafshan:2020fvn,
    author = "Bazrafshan, A. and Olamaei, A. R.",
    title = "{Surface terms of quintic quasitopological gravity and thermodynamics of quasi-topological magnetic brane coupled to nonlinear electrodynamics}",
    eprint = "2012.13828",
    archivePrefix = "arXiv",
    primaryClass = "gr-qc",
    doi = "10.1140/epjc/s10052-022-10250-y",
    journal = "Eur. Phys. J. C",
    volume = "82",
    number = "4",
    pages = "332",
    year = "2022"
}

@article{Olamaei:2023csx,
    author = "Olamaei, A. R. and Bazrafshan, A. and Ghanaatian, M.",
    title = {{Reissner{\textendash}Nordstr{\"o}m black holes in quintic quasi-topological gravity}},
    eprint = "2307.08105",
    archivePrefix = "arXiv",
    primaryClass = "gr-qc",
    doi = "10.1140/epjc/s10052-024-12389-2",
    journal = "Eur. Phys. J. C",
    volume = "84",
    number = "1",
    pages = "46",
    year = "2024"
}

@article{Tsuda:2026xjc,
    author = "Tsuda, Ren and Suzuki, Ryotaku and Tomizawa, Shinya",
    title = "{Fan-Wang type regular black holes in quasitopological gravity}",
    eprint = "2602.16754",
    archivePrefix = "arXiv",
    primaryClass = "gr-qc",
    reportNumber = "TTI-MATHPHYS-39",
    doi = "10.1103/q4kl-6d3z",
    journal = "Phys. Rev. D",
    volume = "114",
    number = "2",
    pages = "024001",
    year = "2026"
}

@article{Frolov:2026rcm,
    author = "Frolov, Valeri P. and Zelnikov, Andrei",
    title = "{Regular black holes in quasitopological gravity: Null shells and mass inflation}",
    eprint = "2601.01861",
    archivePrefix = "arXiv",
    primaryClass = "gr-qc",
    reportNumber = "Alberta Thy 5-25",
    doi = "10.1103/nrd7-j6hs",
    journal = "Phys. Rev. D",
    volume = "113",
    number = "8",
    pages = "084007",
    year = "2026"
}

@article{Hao:2025utc,
	author = "Hao, Chen-Hao and Jing, Jiliang and Wang, Jieci",
	title = "{Charged regular black holes from quasi-topological gravities in D {\ensuremath{\geq}} 5}",
	eprint = "2512.04604",
	archivePrefix = "arXiv",
	primaryClass = "gr-qc",
	doi = "10.1088/1475-7516/2026/05/067",
	journal = "JCAP",
	volume = "05",
	pages = "067",
	year = "2026"
}

@article{Hennigar:2025yqm,
	author = "Hennigar, Robie A. and Kubiz{\v{n}}{\'a}k, David and Murk, Sebastian and Soranidis, Ioannis",
	title = "{Thermodynamics of regular black holes in anti-de Sitter space}",
	eprint = "2505.11623",
	archivePrefix = "arXiv",
	primaryClass = "gr-qc",
	doi = "10.1007/JHEP11(2025)121",
	journal = "JHEP",
	volume = "11",
	pages = "121",
	year = "2025"
}

@article{Bueno:2025zaj,
	author = "Bueno, Pablo and Cano, Pablo A. and Hennigar, Robie A. and Murcia, {\'A}ngel J.",
	title = "{Regular black hole formation in four-dimensional nonpolynomial gravities}",
	eprint = "2509.19016",
	archivePrefix = "arXiv",
	primaryClass = "gr-qc",
	doi = "10.1103/8f3j-zcxh",
	journal = "Phys. Rev. D",
	volume = "113",
	number = "2",
	pages = "024019",
	year = "2026"
}

\end{document}